\documentclass{edm_article}
\usepackage{booktabs}
\usepackage{graphicx}
\usepackage{xcolor}
 \usepackage{url} 
\usepackage{placeins}
\usepackage{float}
\usepackage{comment}
\usepackage[utf8]{inputenc}

\usepackage{amsmath}
\usepackage{hhline}
\usepackage{enumitem}
\usepackage{url}
\usepackage{hyperref}

\begin{document}

\title{Multi-Stage Speaker Diarization for Noisy Classrooms
    \titlenote{(Does NOT produce the permission block, copyright information nor page numbering). For use with edm\_article.cls.}
    }

\numberofauthors{4}
\author{
\alignauthor Ali Sartaz Khan\\
       \affaddr{Stanford University}\\
       \email{askhan1@stanford.edu}
\alignauthor Tol\'{u}l\d{o}p\d{\'{e}} \`{O}g\'{u}nr\d{\`{e}}m\'{i}\\
       \affaddr{Stanford University}\\
       \email{tolulope@stanford.edu}
\and
\alignauthor Ahmed Attia\\
       \affaddr{University of Maryland}\\
       \email{aadel@umd.edu}
\alignauthor Dorottya Demszky\\
       \affaddr{Stanford University}\\
       \email{ddemszky@stanford.edu}
}

\maketitle

\begin{abstract}
Speaker diarization, the process of identifying "who spoke when" in audio recordings, is essential for understanding classroom dynamics. However, classroom settings present distinct challenges, including poor recording quality, high levels of background noise, overlapping speech, and the difficulty of accurately capturing children's voices. This study investigates the effectiveness of multi-stage diarization models using Nvidia’s NeMo diarization pipeline. We assess the impact of denoising on diarization accuracy and compare various voice activity detection (VAD) models, including self-supervised transformer-based frame-wise VAD models. We also explore a hybrid VAD approach that integrates Automatic Speech Recognition (ASR) word-level timestamps with frame-level VAD predictions. We conduct experiments using two datasets from English speaking classrooms to separate teacher vs student speech and to separate all speakers. Our results show that denoising significantly improves the Diarization Error Rate (DER) by reducing the rate of missed speech. Additionally, training on both denoised and noisy datasets leads to substantial performance gains in noisy conditions. The hybrid VAD model leads to further improvements in speech detection, achieving a DER as low as 17\% in teacher-student experiments and 45\% in all-speaker experiments. However, we also identified trade-offs between voice activity detection and speaker confusion. Overall, our study highlights the effectiveness of multi-stage diarization models and integrating ASR-based information for enhancing speaker diarization in noisy classroom environments.

\end{abstract}


\section{Introduction}

Understanding classroom interactions is essential for enhancing student engagement, promoting collaborative learning, and implementing effective teaching strategies. By examining who speaks, when, and for how long, we can gain valuable insights into student participation. Teachers can then use this information to refine their teaching methods and promote collaborative learning, which is crucial for developing students' critical thinking skills \cite{fung2016collaborative,wang2014automatic,demszky2024does}. 

However, capturing accurate speaker information in real-world classroom settings presents substantial challenges. Issues such as a lack of data for children's speech, low signal to noise ratio, speech disfluencies, overlapping speech and multiparty chatter complicate such tasks \cite{southwell2022challenges}. This makes it challenging for conventional audio processing methods like Automatic Speech Recognition (ASR) to transcribe classroom conversations \cite{southwell2022challenges, cao2023comparative} and distinguish between multiple speakers \cite{chang2020endtoendmultispeakerspeechrecognition}. Even though considerable strides have been made to improve the automatic transcription of classroom audio \cite{adel2024classroomw2v2, Park_2020}, transcription alone does not solve the problem of identifying "who spoke when", which is crucial for analyzing individual participation. 

Speaker diarization is a technique that helps solve the ``who spoke when" problem by identifying and labeling different speakers in classroom recordings. 
With the recent advancements in deep learning, speaker diarization has made significant progress over the years \cite{park2021reviewspeakerdiarizationrecent}. State-of-the-art open source models such as NeMo \cite{harper2019nemo} and Pyannote \cite{pyannote} achieve diarization error rates (DER) as low 9\% on multi-party adult conversations. Based on our analyses, the same models, however, perform significantly worse on classroom data off-the-shelf (52-62\% DER). Research remains limited on diarization systems that tackle the challenges of noisy classrooms \cite{wang2024speaker,gomez2022speakerdiarizationidentificationsinglechannel}. While recent efforts have made significant progress using fine-tuning and speaker enrollment \cite{wang2024speaker}, these experiments are challenging to replicate due the privacy of validation data, and difficulty to obtain clean voice samples in many real world classroom environments, which are required for speaker enrollment.

This paper systematically explores \textbf{Nvidia's NeMo multi-stage speaker diarization model} \cite{harper2019nemo}, a cutting edge open-source model that does not require speaker enrollments, and seeks to adapt it to noisy classroom environments. We implement and evaluate the diarization pipeline, incorporate \textbf{denoising techniques}, and \textbf{adapt the Voice Activity Detection (VAD) and speaker embedding models}. Additionally, we introduce a \textbf{hybrid VAD approach} that combines frame-level VAD outputs with word-level timestamps from Whisper's ASR model \cite{radford2022robustspeechrecognitionlargescale}. We apply these models to two datasets: ClassBank \cite{classbank}, an open-source dataset of noisy classroom recordings to enable replicability, and M-Powering Teachers (MPT), a private dataset of higher quality classroom recordings.

 We conduct experiments to separate teacher vs student speech, and to separate all speakers. Our results show that denoising significantly improves DER by reducing the rate of missed speech. Additionally, training on both denoised and noisy datasets leads to substantial performance gains in noisy conditions. Using the hybrid VAD model, we observed further improvements in speech detection, \textbf{achieving a DER as low as 17\% in teacher-student experiments and 45\% in all-speaker experiments}.  We also identified trade-offs between VAD performance and speaker confusion. 
 Our qualitative error analyses indicate that,  shorter speech segments from students contribute the most to diarization errors, whereas errors are more uniform across segment durations for teachers. To facilitate further research in this domain, we released our code publicly.\footnotemark \footnotetext{\raggedright Code available at \url{https://github.com/EduNLP/nemo-multistage-classroom-diarization}}

 
 


\section{Related Work}

\subsection{Speech Recognition in Classroom Settings}
Prior studies have highlighted challenges in achieving accurate transcriptions in noisy classrooms. Both Southwell et al. \cite{southwell2022challenges} and Cao el al. \cite{cao2023comparative} evaluated multiple ASR engines for transcribing recordings of student classroom discourse and found high Word Error Rates (WER) across all systems. Similarly, Dutta et al. \cite{dutta-etal-2022-activity} focused on preschool environments and achieved WERs ranging from 28\% to 47\% across different classroom activities. Kelly et al. \cite{kelly2018automatically} used ASR to identify authentic questions in the classrooms to support better teaching initiatives. Sun et al. \cite{sun2024said} proposed a new automated framework that achieves a 15\% WER in preschool classrooms using a combination of ALICE \cite{rasanen2021alice} for speaker classification and Whisper \cite{whisper} for transcription. Our paper introduces a hybrid approach that integrates frame-level VAD and ASR outputs to enhance VAD performance, showcasing a use of ASR outputs that will further benefit from advancements in ASR models.

\subsection{Speaker Diarization in Classroom Settings}
Shifting to diarization, Wang et al. \cite{wang2024speaker} developed a system combining ECAPA-TDNN \cite{desplanques2020ecapa} embeddings with Whisper ASR and speaker enrollments to achieve a DER of 34.46\% in noisy classrooms. Gómez et al. \cite{gomez2022speakerdiarizationidentificationsinglechannel} introduced virtual microphone arrays to enhance speaker identification in group discussions and outperformed services like Google Cloud \cite{GoogleCloud2021} and Amazon AWS \cite{AmazonTranscribe2021}. Dubey et al. \cite{dubey2016speaker} proposed an unsupervised system for peer-led team Learning sessions which utilized online speaker change detection and Hausdorff-distance-based clustering to improve DER. Moreover, Cánovas and García Clemente \cite{canovas2022analysis} integrated speaker diarization with non-verbal discourse feature extraction (such as participant speaking ratio, average pause duration, etc) to analyze classroom interactions and distinguish between teachers and students. Similar to Wang et al. \cite{wang2024speaker}, this paper utilizes a multi-stage speaker diarization pipeline in an effort to improve DER in noisy classrooms. However, due to the lack of speaker enrollments in our datasets, we propose alternative methods that do not rely on clean voice samples.

\subsection{Novel Approaches to Speaker Diarization}
Recent advances in speaker diarization have introduced several innovative methods. Fujita et al.'s \cite{fujita2020neural} SC-EEND model improves DER for multi-speaker scenarios using a speaker-wise chain rule. Gomez et al.’s \cite{gomez2022speakerdiarizationidentificationsinglechannel} virtual microphone approach enhances diarization in noisy classrooms without extensive training data. Kanda et al. \cite{kanda2022transcribe} introduced the Transcribe-to-Diarize method that enables end-to-end speaker-attributed ASR for an unlimited number of speakers. Since research into diarization in classrooms remains heavily limited, we begin with evaluating and adapting simpler and widely used open source approaches. 

\section{Data}
\label{data}
This study utilized two datasets: ClassBank \cite{classbank} and the M-Powering Teachers (MPT) Data \cite{adel2024classroomw2v2}. These datasets provide $\sim$60 hours of speech recordings that reflect various classroom environments, speaker demographics, and noise conditions.

\subsection{ClassBank Dataset}
The ClassBank dataset \cite{classbank} is a publicly available dataset that provides access to transcribed filmed interactions in a variety of classrooms. Topics include science, mathematics, medicine, and reading. It captures interactions among students and instructors starting from third grade all the way to medical school. We filtered the dataset for English classroom recordings and for fully-annotated files resulting in 44 hours of audio recordings (322 files). We divided the files into a 33-hour training set (257 files), a 5.5-hour development set (32 files), and a 5.5-hour test set (33 files). Each recording averages 10 minutes in duration, with 1 to 19 speakers per session.  The data includes manual transcriptions with speaker-segment level timestamps. We selected this data for its public availability and representation of various classroom settings, which makes it a valuable resource for research on classroom ASR and diarization.

\subsection{M-Powering Teachers (MPT) Dataset}

The MPT Dataset \cite{adel2024classroomw2v2} is a private dataset that comprises 6 hours of classroom audio (12 files) where the teacher and students interact with each other. The recordings come from six 5-8th grade math classrooms across three schools in California, Ohio, and Washington D.C.  The selected schools span private, charter, and public schools that serve different communities including White, Asian, and African-American students from various socio-economic backgrounds. This data includes manual transcriptions with speaker-segment level timestamps. We split the recordings into a 3.5-hour training set (7 files), a 1-hour development set (1 file), and a 1.5-hour test set (4 files). The test set includes one full-hour recording and three ten minute segments sampled to represent various classroom engagement phases across multiple transcripts. We ensured there was no data leakage by excluding those segments from the training set. In contrast to ClassBank, these recordings have a longer average duration of 48 minutes per session, featuring 11 to 18 speakers. As the audio was recorded with high-quality equipment to produce data for classroom transcript analysis, the data contains lower noise levels than ClassBank. 



\subsection{Denoising Audio}
\label{subsec:denoising_audio}
\FloatBarrier 
\begin{figure}[h]
    \centering
\includegraphics[width=0.45\textwidth]{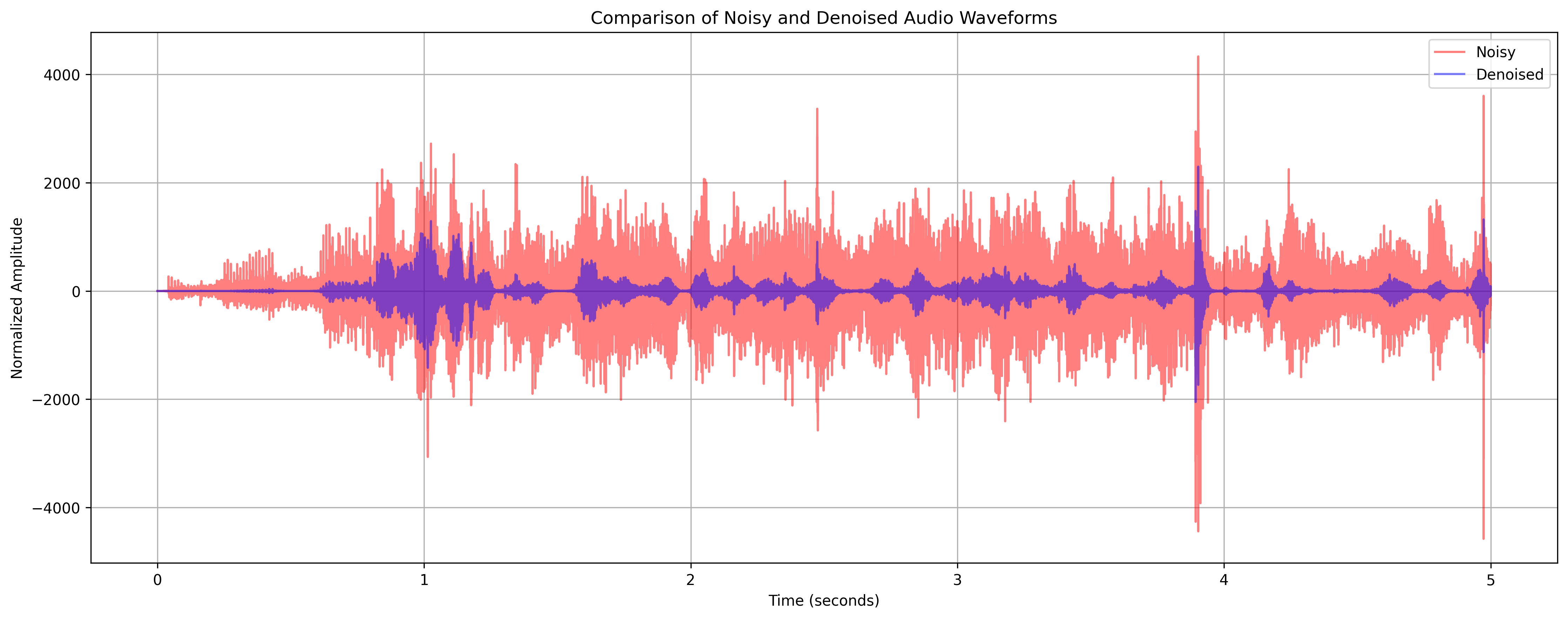}
\Description{Noisy (red) vs. Denoised (purple) speech segment comparison}

    \caption{Noisy (red) vs. Denoised (purple) speech segment comparison}
    \label{fig:noisyvsdenoised}
\end{figure}

Removing noise from audio is a difficult task, leading to several possible approaches \cite{denoising1_yu, denoising2_zhou, denoising3_wung, sainburg2020finding}. We employed Sainburg et al.'s  noise reduction algorithm \cite{tim_sainburg_2019_3243139} for audio preprocessing. This algorithm effectively minimizes background noise (see Figure \ref{fig:noisyvsdenoised}) and improves overall speech clarity. We first experimented with denoising audio at inference time, to estimate its effects on DER. We observed that denoising inadvertently suppressed children's speech segments due to the nature of the audio recording setup. Since the microphone was positioned closer to the teacher, the voices of many students, particularly those seated at the back of the classroom, were recorded at significantly lower volumes. As a result, during denoising, these low-amplitude speech signals were often removed, which led to the unintended loss of student contributions. Thus, instead of applying denoising at inference time, which showed poor results after training models, we used this technique as a data augmentation tool: we included both the original and denoised versions of each audio recording to create a more robust training dataset.  Our final training set included approximately 70 hours of audio, which combined recordings from the ClassBank and MPT dataset training splits along with their denoised counterparts.

\section{Diarization Architecture}

\begin{figure}[t]
\FloatBarrier
    \centering
    \includegraphics[width=0.45\textwidth]{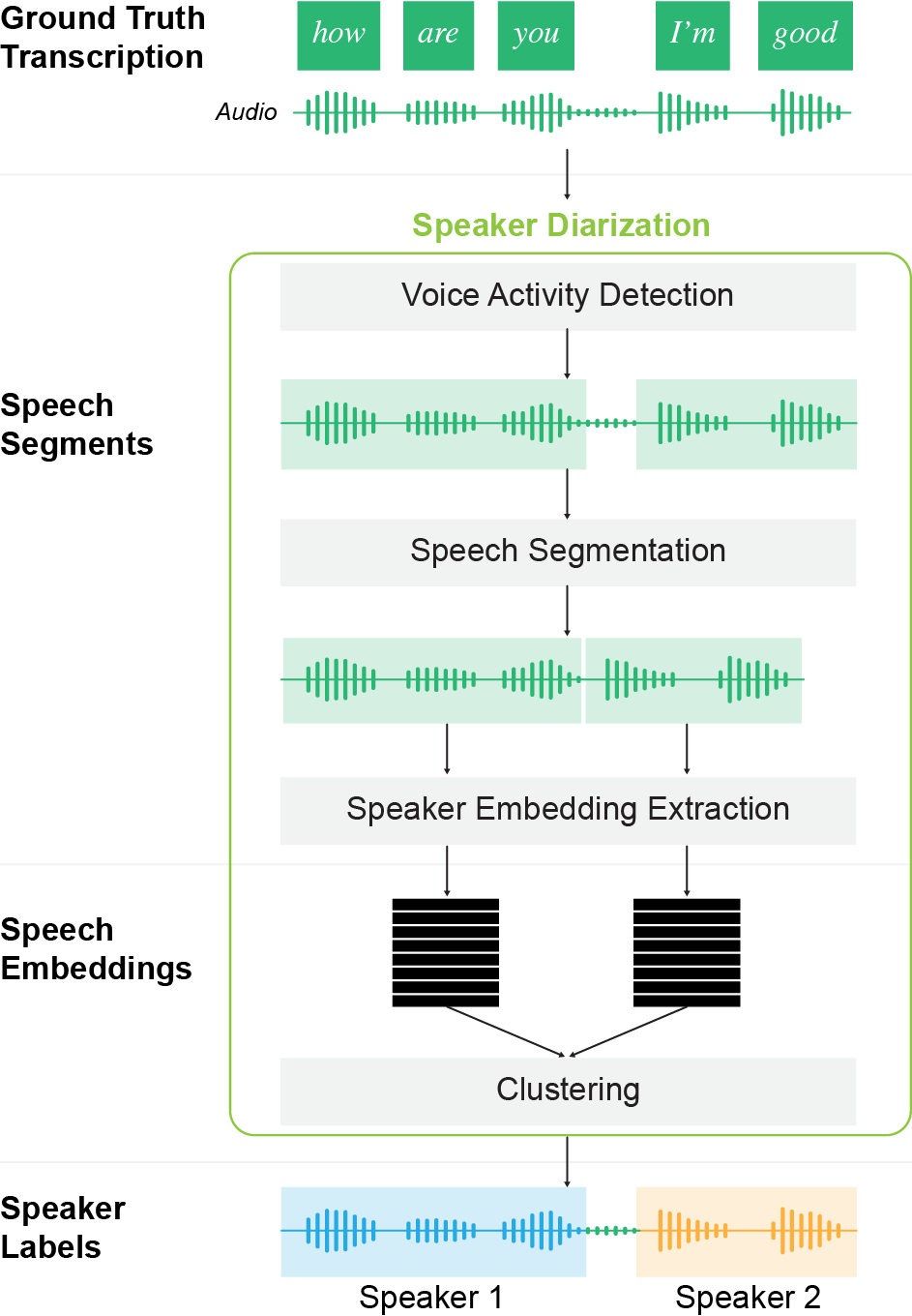}
    \Description{NeMo Multi-Stage Diarization Architecture}
    \caption{NeMo Multi-Stage Diarization Architecture}
    \label{fig:nemo_architecture}
\end{figure}

Our model architecture customizes \textbf{NeMo’s multi-stage diarization pipeline} \cite{NemoDiarization2024}. Given an input audio file, this pipeline generates timestamps for detected speech segments and assigns a speaker label to each segment. Segments are defined as continuous audio clips containing speech that are defined by start and end timestamps. The pipeline is structured into three key modules, as illustrated in Figure~\ref{fig:nemo_architecture}: Voice Activity Detection, Speaker Embedding Extraction, and Clustering. Unlike end-to-end models, this multi-stage approach allows us to optimize individual components, identify which stage contributes most to errors, and swap in custom models to improve performance.

\subsection{Voice Activity Detection}

Voice Activity Detection (VAD) identifies where speech occurs in the audio. We evaluated four VAD approaches: segment-wise VAD, frame-wise VAD, speech recognition models as VAD, and a hybrid method combining ASR and frame-level VAD outputs. We measure performance using missed detection rate (\textbf{MISS}) and false alarm rate (\textbf{FA}). Missed detection occurs when the VAD fails to detect a speech segment, whereas false alarm happens when non-speech regions are mistakenly classified as speech.

\subsubsection{Segment-wise VAD} \label{subsec:segment_vad} Segment-wise VAD predicts whether an entire audio segment contains speech. First, the audio is split into overlapping segments (0.63s segments with 0.01s shift). Each segment is fed into the VAD model for a speech probability, then the overlapping predictions merge to yield frame-level predictions. We used NeMo’s default segment-wise models: VAD Multilingual MarbleNet \cite{vad_multilingual_marblenet}, VAD MarbleNet \cite{vad_marblenet}, and VAD Telephony MarbleNet \cite{vad_telephony_marblenet}, which are CNN-based. VAD Multilingual MarbleNet performed best, so we fine-tuned it on our training data to tailor it to classroom audio.

\subsubsection{Frame-wise VAD using Wav2vec} \label{subsec:frame_vad} Frame-wise VAD generates a speech probability for each 20ms frame. We used NeMo’s default Frame-VAD Multilingual MarbleNet \cite{vad_frame_marblenet} and also trained a custom VAD with \textbf{Wav2vec 2.0} (w2v2) \cite{baevski2020wav2vec}. We added a linear classification head to w2v2 for predicting speech frames, experimenting with two variants: \textbf{Robust-Large} \cite{w2v2_robust_large},  with 300 million parameters designed for robustness in noisy environments, and \textbf{CPT-Boosted w2v2} \cite{adel2024classroomw2v2}, a model that was further pre-trained on noisy classroom recordings. Each variant was tested under two training schemes: (1) training only the classifier and (2) training both w2v2 and the classifier. We labeled frames via ground-truth Rich Transcription Time Marked (RTTM) files (20ms per frame) and used identical training configurations (10 epochs, learning rate 1e-4). During training, we used 2s windows with a 0.25s stride in frames to capture temporal dependencies while maintaining fine-grained resolution.  We selected checkpoints based on the best F1 on the validation set. Training both the base model and classification head generally yielded the highest performance, though gains were marginal over training only the classifier.

\subsubsection{Speech Recognition Models as VAD} \label{subsec:whisper} Whisper \cite{whisper} is an open-source ASR model that provides word-level timestamps. We converted these to frame-level VAD outputs by labeling all frames within detected word timestamps as speech. We used Whisper’s large-v2 model, which is pre-trained on 680k hours of labelled data, and large-v3 model, which was trained on 1 million hours of weakly labeled audio plus 4 million hours of pseudo-labeled audio.

\subsubsection{Hybrid VAD: Combining Frame-wise VAD and Whisper} \label{subsec:hybrid_vad}

Frame-wise VAD yielded low MISS but high FA, while Whisper had lower FA but higher MISS. We therefore combined their frame-level outputs into a hybrid model. For each frame $i$, we computed a weighted sum of the predictions from both models using the following formula: 
\begin{equation}
    \text{Y}_i = \alpha \cdot \text{frame-vad}_i + (1 - \alpha) \cdot \text{whisper}_i 
    \end{equation}
The parameter \(\alpha\) controlled the balance between the two models, allowing us to fine-tune the influence of each VAD system. 

\subsubsection{VAD Thresholds} \label{subsec:vad_thresholds} Thresholds determine how a model classifies speech vs. non-speech frames. We tuned the \textbf{onset} threshold (probability to classify a frame as speech), the \textbf{offset} threshold (probability below which a previously classified speech frame is reclassified as non-speech), and the weighting parameter $\boldsymbol{\alpha}$. We tested onset values from 0.3--0.9, offset from 0.1--0.8, in increments of 0.05, and identified the best performing combinations with and without Whisper integration. Combining frame-wise VAD with Whisper-derived timestamps reduced both FA and MISS.

\subsection{Speaker Embedding Extraction} \label{subsec:speaker_emb} Speaker embedding extraction converts VAD-detected speech segments into fixed-length vectors, (\textbf{speaker embeddings}), capturing speaker-specific characteristics. These embeddings are used in clustering, with better embeddings yielding fewer speaker confusions. We tested three NeMo speaker embedding models: TitaNet-Large \cite{NemoTitaNetLarge2024}, ECAPA-TDNN \cite{NemoECAPATDNN2024}, and SpeakerNet \cite{NemoSpeakerNet2024}. TitaNet-Large performed best; however, fine-tuning it on our data degraded performance (perhaps due to overfitting to a small sample), so we used the pre-trained version.

\begin{figure}[h]
    \centering
    \includegraphics[width=0.9\linewidth]{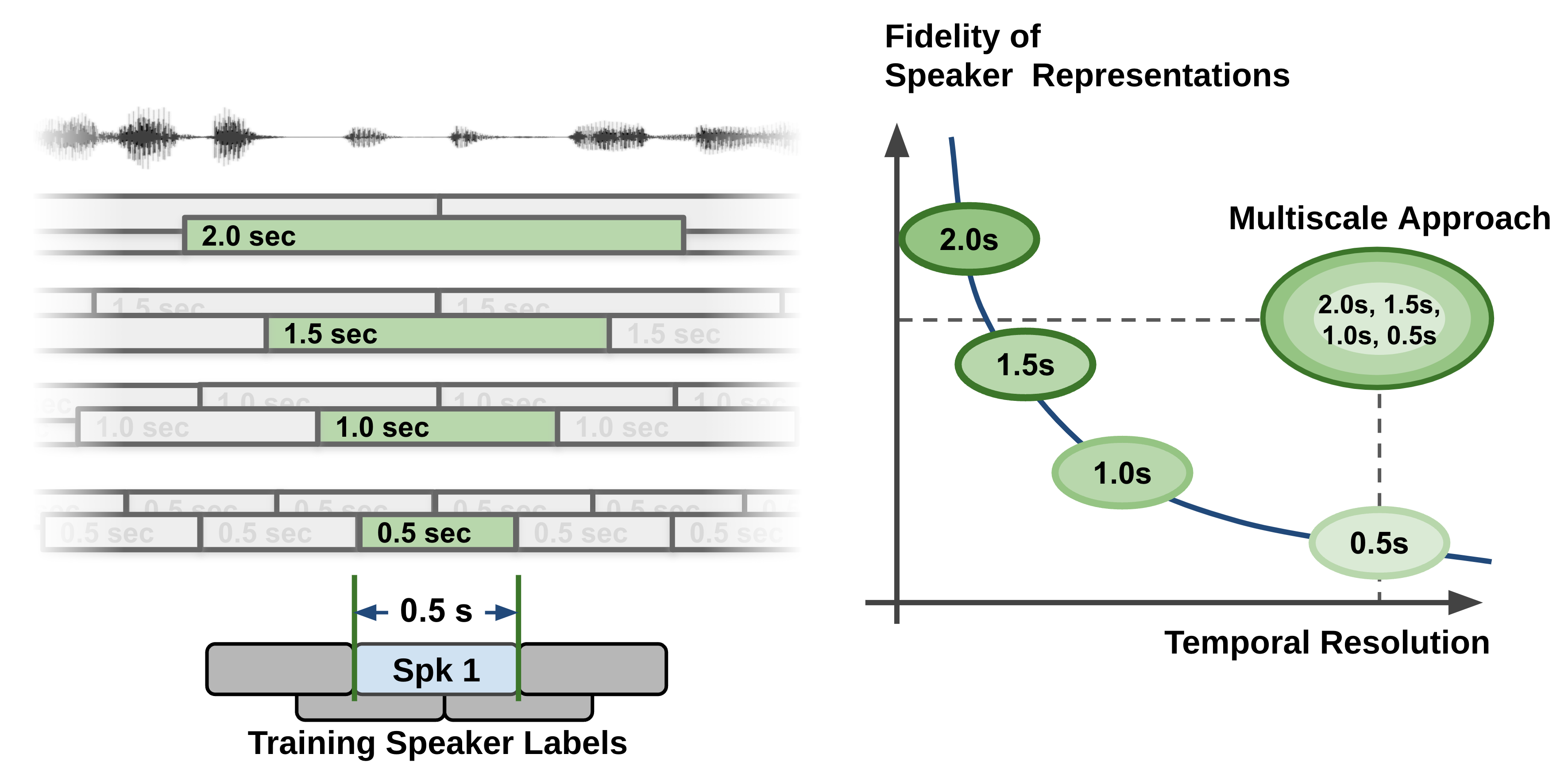}
    \caption{Illustration of NeMo's multi-scale segmentation approach for speaker embedding extraction 
    \cite{NemoDiarization2024}.}
    \Description{Illustration of NeMo's multi-scale segmentation approach for speaker embedding extraction 
    \cite{NemoDiarization2024}.}
    \label{fig:speaker_embedding}
\end{figure}

Embedding extraction balances \textbf{speaker quality} and \textbf{timestamp granularity}: longer segments improve speaker representations but offer coarser labeling. Traditional systems often use segments of 1.5--3.0s, but this reduces precision in speaker counting. NeMo addresses this by using a \textbf{multi-scale segmentation} strategy (Figure~\ref{fig:speaker_embedding}), extracting embeddings at multiple segment lengths and combining them for better accuracy \cite{NemoDiarization2024}. This approach improves performance in short utterances, which is particularly useful in classroom environments where students especially talk for short periods.

\subsection{Clustering}

Clustering is the final step in the diarization pipeline, which groups speaker embeddings based on speaker identity. It applies spectral clustering on embeddings extracted from speech segments and assigns labels by similarity. 

This process can either predict the number of speakers automatically or use a predefined (oracle) number of speakers. We found that automatic speaker prediction often led to imbalanced clusters, with the model favoring dominant speakers while underrepresenting others. To ensure more accurate speaker assignments, we used the oracle number of speakers for clustering. So, for our teacher-student experiments, number of speakers was set to 2, and for all-speaker experiments, the number of speakers was set to the groundtruth number of speakers for that respective recording.

\section{Experiments and Results}

We conducted experiments to separate teacher-student speech and separate speech from all speakers. To evaluate diarization performance, we use diarization error rate (DER), which is calculated as the sum of false alarm (FA), missed detection (MISS), and confusion error rate (CER): 
\begin{equation}
DER = (FA + MISS + CER)/{\text{Total Speech Duration}}
\end{equation}

\subsection{Effects of Denoising on DER}
\label{subsec:effects_of_denoising}
We first evaluated the out-of-the-box (ootb) configuration of NeMo. As mentioned in \ref{subsec:segment_vad} and \ref{subsec:speaker_emb}, we used VAD Multilingual MarbleNet \cite{vad_multilingual_marblenet} for voice activity detection (VAD) and TitaNet-Large \cite{NemoTitaNetLarge2024} for speaker embedding extraction. To assess the impact of denoising, we ran tests on both noisy and denoised datasets.

\begin{table}[h]
\centering
\scriptsize
\begin{tabular}{l | c c | c c c c}
\toprule
\multicolumn{7}{c}{\textbf{ClassBank}} \\ 
\midrule
Speakers & Audio & Model & DER & FA & MISS & CER \\
\midrule
Two  & Noisy    & NeMo  & 50.5 & 7.0 & 39.4 & 4.2 \\
     & Denoised & NeMo  & \textbf{36.7} & \textbf{13.9} & \textbf{11.4} & \textbf{11.4} \\
\midrule
All  & Noisy    & NeMo  & 64.6 & 7.0 & 39.4 & 18.3 \\
     & Denoised & NeMo  & \textbf{58.1} & \textbf{13.9} & \textbf{11.4} & \textbf{32.8} \\
\bottomrule
\end{tabular}


\begin{tabular}{l | c c | c c c c}
\multicolumn{7}{c}{\textbf{MPT}} \\ 
\midrule
Speakers & Audio & Model & DER & FA & MISS & CER \\
\midrule
Two  & Noisy    & NeMo  & 33.2 & 1.1 & 29.6 & 2.5 \\
     & Denoised & NeMo  & \textbf{26.8} & \textbf{8.8} & \textbf{11.7} & \textbf{6.4} \\
\midrule
All  & Noisy    & NeMo  & \textbf{71.3} & \textbf{1.1} & \textbf{29.6} & \textbf{40.6} \\
     & Denoised & NeMo  & 82.2 & 8.8 & 11.7 & 61.7 \\


\bottomrule
\end{tabular}

\caption{Evaluating the utility of inference-time denoising using NeMo out-of-the-box model.}
\label{table:denoised_results}
\end{table}

Unsurprisingly, the teacher-student setup outperformed the all-speaker setup (see Table \ref{table:denoised_results}) because of the disparity in number of speakers. Denoising improved DER more for ClassBank than the MPT dataset, likely because ClassBank had noisier recordings that benefited more from noise reduction. While denoising significantly reduced MISS, it also increased FA which suggests that some noise was misclassified as speech. The confusion errors were higher in denoised datasets compared to their noisy counterparts. When our diarization system detects more speech, it provides the speaker embedding model with more data to process. However, it also increases the chances of incorrect speaker assignments. This highlights a trade-off, that we notice throughout all our experiments, between better speech detection and accurate speaker assignments. Building on these findings, we augment the training data with denoised data for all downstream experiments, but do not use denoising at inference time to mitigate speaker confusion errors.

\subsection{Effects of Data Augmentation on DER}

As mentioned in Section \ref{subsec:denoising_audio}, rather than applying denoising at inference time, we used it as a data augmentation strategy. We included both the original and denoised versions of each audio recording to create a more robust training set. We then finetuned NeMo's VAD Multilingual MarbleNet model \cite{vad_multilingual_marblenet} with and without this augmentation. We observed improvements in DER for both the ClassBank and MPT datasets for the teacher-student experiments, with MPT showing a particularly notable improvements (see Table \ref{tab:nemo_data_aug_results}). However, performance slightly degraded in the all-speaker experiments. Since our error analysis in Section \ref{sec:qual_errors} focused on the Teacher-Student experiments, we chose to proceed with data augmentation in all subsequent experiments.

\begin{table}[h]
\centering
\scriptsize
\begin{tabular}{l c c c}
\toprule
\textbf{Dataset} & \textbf{Data Augmentation} & \textbf{TS DER} &  \textbf{AS DER}\\
\midrule
ClassBank & Yes & \textbf{33.1} & 51.2 \\
          & No & 33.2 & \textbf{49.5} \\
\midrule
MPT       & Yes & \textbf{22.5} & 74.4 \\
          & No & 26.7 & \textbf{73.5} \\
\bottomrule  
\end{tabular}
\caption{Diarization results from finetuning the VAD Multilingual MarbleNet model with and without data augmentation evaluated on Teacher-Student (TS) and All-Speaker (AS) experiments.}
\label{tab:nemo_data_aug_results}
\end{table}

\subsection{Comparing state-of-the-art model performances in classrooms}
We tested NeMo's diarization model \cite{NemoDiarization2024} and Pyannote \cite{pyannote} on our classroom datasets (see Table \ref{tab:nemo_pyannote_results}). These state-of-the-art (SOTA) diarization models serve as strong baselines. For NeMo, we used the best results from our denoising experiments, while Pyannote showed no noticeable improvement after data augmentation with denoising. This suggests that, unlike multi-stage diarization models, neural diarization models might not benefit significantly from pre-processing techniques like denoising. This is likely because end-to-end models learn noisy features directly from raw audio, which reduces the need for external denoising. We decided to prioritize NeMo because it showed better performance on teacher-student experiments, which allowed for more comprehensive error analysis on teacher-student classroom dynamics.

\begin{table}[h]
\centering
\scriptsize
\begin{tabular}{l c c c}
\toprule
\textbf{Dataset} & \textbf{Model} & \textbf{TS DER} &  \textbf{AS DER}\\
\midrule
ClassBank & NeMo & 36.7 & 58.1 \\
          & Pyannote & 45.4 & 60.1 \\
\midrule
MPT       & NeMo & 26.8 & 71.3 \\
          & Pyannote & 32.9 & 54.6\\
\bottomrule  
\end{tabular}
\caption{Baseline diarization results for out-of-the-box NeMo and Pyannote models on Teacher-Student (TS) and All-Speaker (AS) experiments.}
\label{tab:nemo_pyannote_results}
\end{table}

\subsection{Effects of VAD models on DER}
Table~\ref{table:nemo_diarization_results} presents the best-performing VAD configurations. As our baseline, we use NeMo's best performing VAD model, VAD Multilingual MarbleNet \cite{vad_multilingual_marblenet}, after hyperparameter tuning. 
Similarly to the denoising experiments, we use NeMo's Titanet-Large \cite{NemoTitaNetLarge2024} as the primary speaker embedding model in all experiments below.

\subsubsection{Impacts of Finetuning Baseline VAD}
As mentioned in \ref{subsec:denoising_audio}, we fine-tuned VAD Multilingual Marblenet on our training set that contained both denoised and noisy audio recordings. After running evaluations on our noisy test set, the fine-tuned model consistently performs equal to or better than the tuned Nemo VAD (Table~\ref{table:nemo_diarization_results}) by reducing missed detection rates while keeping false alarms low. This makes the fine-tuned model a more reliable choice for diarization in both two-speaker and all-speaker settings.

\subsubsection{Impacts of Continued Pre-Training (CPT)}
As mentioned in \ref{subsec:frame_vad}, we trained custom VAD models using two variants of Wav2vec 2.0 (w2v2) \cite{baevski2020wav2vec}: Robust-Large \cite{w2v2_robust_large} and CPT-Boosted on classroom data \cite{adel2024classroomw2v2}. As shown in Table~\ref{table:nemo_diarization_results}, the CPT-Boosted w2v2 VAD marginally outperforms the Robust-Large w2v2 VAD on the ClassBank dataset \cite{classbank} in our two-speaker experiments. However, the Robust-Large w2v2 VAD outperforms the CPT-Boosted model at almost every other experiment, including two-speaker and all-speaker experiments using the MPT \cite{adel2024classroomw2v2} and even all-speaker experiments using the ClassBank dataset. This shows us that while CPT-Boosted w2v2 models perform well in classroom speech recognition tasks \cite{adel2024classroomw2v2}, those performance gains don't translate to improved VAD performance.

\subsubsection{Impacts of incorporating ASR timestamps}
As mentioned in \ref{subsec:whisper}, we combined word-level timestamps from Whisper ASR with frame-wise VAD outputs to balance the complementary advantages of each approach. We used Whisper large-v3 for the ClassBank dataset and large-v2 for the MPT dataset, because these were the best-performing models for each respective dataset. We use this hybrid VAD approach on two of our best-performing models on each dataset. We see DER improvements of up to \textbf{5.6\%} using this hybrid VAD method. The primary reasons for this performance boost are the model's improved ability to detect speech segments while minimizing false alarms. As mentioned in \ref{subsec:effects_of_denoising}, we noticed confusion error rates increase when VAD performance improves, and vice-versa. As a result, we had to find a balance between the two metrics to obtain the best DER results. In the future, we might expect more drastic improvements in DER with better performing ASR and VAD models tailored for noisy classrooms.

\subsubsection{Trade-offs of different VAD methods}
In this work, we employ several tactics using differing model architectures to do VAD. The performance varies, with the addition of leveraging Whisper timestamps improving performance, the additional computational overhead could be limiting in some cases. Finetuning the VAD Multilingual MarbleNet model is sure to improve performance and if computation allows, incorporating Whisper will surely improve results.


\begin{table}[h]
\centering
\scriptsize
\begin{tabular}{l | c | c}
\toprule
\multicolumn{3}{c}{\textbf{ClassBank}} \\ 
\midrule
\textbf{Model} & \textbf{TS DER} & \textbf{AS DER} \\
\midrule
NeMo &  36.7 & 58.1 \\
+ Fine-tuned Multilingual Marblenet & 30.5 & 48.9 \\
\hspace{6pt} + Whisper & 29.9 & \textbf{45.4} \\
+ CPT-Boosted w2v2 & 30.8 & 49.1 \\
\hspace{6pt} + Whisper & \textbf{29.6} & 46.5 \\
\midrule
+ Tuned Multilingual Marblenet & 31.0 & 49.1 \\
+ Robust-Large w2v2 & 31.6 & 48.4\\
+ Whisper VAD & 36.3 & 50.8 \\
+ Frame-VAD Multilingual Marblenet & 42.7 & 63.4\\
\bottomrule
\end{tabular}


\begin{tabular}{l | c | c}
\multicolumn{3}{c}{\textbf{MPT}} \\ 
\midrule

\textbf{Model} & \textbf{TS DER} & \textbf{AS DER} \\
\midrule
NeMo & 26.8 & 71.3  \\
+ Robust-Large w2v2 & 21.7 & 50.0 \\
\hspace{6pt} + Whisper & \textbf{17.4} & \textbf{44.4} \\
+ Fine-tuned Multilingual Marblenet & 21.0 & 47.7 \\
\hspace{6pt} + Whisper & 18.1 & 46.3 \\
\midrule
+ Tuned Multilingual Marblenet & 24.4 & 74.1 \\
+ Whisper VAD & 24.3 & 64.2 \\
+ CPT-Boosted w2v2 & 26.2 & 68.9 \\
+ Frame-VAD Multilingual Marblenet & 30.4 & 64.2\\


\bottomrule
\end{tabular}

\caption{Diarization results for Teacher-Student (TS) and All-Speaker (AS) experiments using variations of the NeMo diarization model with different VAD configurations. The table presents diarization error rates (DER) for models tested on the ClassBank and MPT datasets, showcasing the impact of hyperparameter-tuning, fine-tuning, and VAD selection.}
\label{table:nemo_diarization_results}
\end{table}

\begin{table}[h]
\centering
\scriptsize
\begin{tabular}{l | c c c | c c c}
\toprule

\textbf{ClassBank} & \multicolumn{3}{c|}{\textbf{Student}} & \multicolumn{3}{c}{\textbf{Teacher}} \\
\midrule
Type & Long & Medium & Short & Long & Medium & Short \\ 
\midrule
CER   & 4.3  & 7.1   & 14.8  & 16.5  & 10.5  & 14.7  \\
FA    & 2.1  & 2.1   & 3.1   & 2.2   & 1.4   & 6.5   \\
MISS  & 0.1  & 0.0   & 0.6   & 0.0   & 0.0   & 0.5   \\
Correct & 93.5  & 90.8  & 81.6  & 81.3  & 88.1  & 78.3  \\
\midrule
Ratio & 14.3  & 26.8  & 58.9  & 22.3  & 35.2  & 42.5  \\

\end{tabular}

\begin{tabular}{l | c c c | c c c}
\toprule
\textbf{MPT} & \multicolumn{3}{c|}{\textbf{Student}} & \multicolumn{3}{c}{\textbf{Teacher}} \\
\midrule
Type & Long & Medium & Short & Long & Medium & Short \\ 
\midrule
CER & 3.6 &  7.2 & 46.3 &  1.2 &  1.9 & 11.1 \\
FA & 8.9 & 14.5 & 17.3 &  2.9 & 20.2 & 40.1 \\
MISS & 0.0 &  1.4 &  1.4 &  0.0 &  0.0 &  0.0 \\
Correct \hspace{4pt} & 87.5 & 76.8 & 34.9 & 95.9 & 77.9 & 48.8 \\
\midrule
Ratio & 11.7 & 14.5 & 73.8 & 39.0 & 23.9 & 37.2 \\
\bottomrule
\end{tabular}
\caption{Error distributions for teacher-student experiments by speaker role and segment duration across ClassBank and MPT datasets. Segments are categorized as short (<2s), medium (2-5s), or long (>5s). For each category, the table presents the percentage of Confusion Error Rate (CER), False Alarm (FA), Missed Detection (MISS), and Correct classifications -- the percentages within each column sum to 100\%. The Ratio row indicates the distribution of each duration category within each speaker role in each dataset (row sums to 100).
}
\label{tab:teacher-student_errors}
\end{table}


\subsection{Qualitative Error Analysis} \label{sec:qual_errors}
\paragraph{Relationships between diarization error metrics}
We fo-und that DER has a strong correlation with VAD performance (calculated using the sum of FA+MISS) both within teacher-student experiments (Spearman $\rho$= 0.96, p < 0.001) and within all speaker experiments ($\rho=0.94$, p < 0.001). The strong correlation between VAD and DER underscores the importance of improving VAD for enhancing diarization performance.  In contrast, CER has a weak correlation with DER in both setups (TS: $\rho$=-0.08, p = 0.83; AS: $\rho=$-0.12, p = 0.73), likely because we did not vary the speaker embedding model and thus its performance within each setup is largely contingent on the outputs of VAD. When we look across experiments, the important contribution of CER becomes evident: the mean CER increases considerably from two-speaker to all-speaker experiments (9.2 vs.\ 29.8), which in-turn also increases mean DER from 32.3 to 53.3. This highlights the need to improve speaker embedding models, especially when distinguishing more than two speakers in classroom contexts.

\paragraph{Relationship between segment duration and diarization errors} The data from Table \ref{tab:teacher-student_errors} demonstrates clear relationships between segment duration and diarization accuracy for both student and teacher speech. For student speech, diarization accuracy decreases consistently as segments become shorter. For MPT, this drop is especially stark: while 87.5\% of long segments are assigned correctly, only 34.9\% of short segments are. This reduced performance for student speech is largely due to an increase in confusion errors (CER) for short segments vs long segments (14.8\% vs 4.3\% in ClassBank, 46.3\% vs 3.6\% in MPT). False alarms (FA) remain relatively stable across segment lengths for students (2.1-3.1\%) in ClassBank, but increases notably for MPT as segments become shorter. Missed detections (MISS) are notably low across all categories for both teachers in students in both datasets $\le 1.4$, suggesting that the main challenges lie in speaker confusion and false alarms rather than failure to detect speech. 

In MPT, teacher speech exhibits a similar pattern to student speech (although with better overall performance): CER and FA rates increase consistently as segment duration gets shorter. The jump in FA is particularly drastic for teachers in MPT, from 2.9\% in long segments to 40.1\% in short segments. However, ClassBank shows a slightly different trend, with medium-length segments achieving higher accuracy (88.1\%) than long segments (81.3\%). This is due to both CER (10.5\%) and FA (1.4\%) being lowest in medium teacher segments. 

The ratio distribution indicates that short segments constitute a substantial portion of both datasets (58.9\% for students, 42.5\% for teachers in ClassBank; 73.8\% for students, 37.2\% for teachers in MPT). This analysis overall highlights the importance of improving diarization performance for short segments, especially for children speech.

\section{Conclusion and Future Work}
Speaker diarization systems play a crucial role in analyzing classroom interactions by distinguishing individual speakers and teacher-student dynamics. This technology offers valuable insights into communication patterns, student engagement, and instructional effectiveness, enabling educators to refine teaching strategies for more interactive discussions.

This paper demonstrates the effectiveness of multi-stage diarization models in addressing challenges in noisy classroom environments. Denoising reduces missed speech detections, while training on both denoised and noisy datasets enhances robustness. The hybrid VAD model, which combines ASR-based timestamps with frame-wise VAD outputs, achieves Diarization Error Rates (DER) as low as 17\% on separating teachers vs student speech, showing performance gains of 9\% compared to a strong baseline model. This approach introduces the potential of integrating ASR outputs with VAD models for improved performance in speech detection.

Despite these advancements, challenges 
still remain in correctly identifying speech segments and assigning the correct speakers to each segment. Future work should focus on refining speaker embedding models and experimenting with end-to-end neural-diarization models such as Pyannote \cite{pyannote} and \cite{fujita2020neural}. Integrating language information into diarization pipelines (e.g. via an integrated ASR+diarization system) and training on additional data from noisy classrooms may further improve the performance of diarization models in real-world classrooms. Additionally, there should be an emphasis on enhancing the detection of short speech segments, particularly for students, as our analysis identifies this as a key limitation in existing classroom diarization systems.

\section{Practical implications} The hybrid pipeline results in speaker‑attributed transcripts which may be accurate enough to support downstream analytics such as the calculation of teacher‑student talk ratios. These analytics enable tools like TeachFX\footnote{\url{www.teachfx.com}}, TalkMoves \cite{suresh2021using} or M‑Powering Teachers \cite{demszky2023m} to surface insights to teachers or coaches without costly manual annotation. The fact that our pipeline can be hosted ``in house'' on a tool's server also helps protect teacher and student privacy. However, the separation of all speakers (i.e. student voices) still has significant room for improvement. As diarization accuracy continues to improve—particularly for short student turns—these classroom analytics can expand to finer‑grained tasks such as identifying brief student responses to teacher questions, and separating student voices and monitoring balance in group work contexts.

\section{Limitations}
Our data is drawn from English-speaking classrooms, is a relatively small sample drawn from particular teacher and student populations, and was recorded often with high-quality equipment. The generalizability of our findings to other classroom data contexts remains to be tested. Our data also lacks voice enrollment information, which makes the diarization task significantly more challenging. Different methods may perform better on data where speaker voice samples are available. 

\bibliographystyle{abbrv}
\bibliography{sigproc}

\begin{thebibliography}{10}

\bibitem{AmazonTranscribe2021}
{Amazon Web Services}.
\newblock Amazon transcribe, 2021.
\newblock \url{https://docs.aws.amazon.com/transcribe/}.

\bibitem{adel2024classroomw2v2}
A.~A. Attia, D.~Demszky, T.~Ogunremi, J.~Liu, and C.~Espy-Wilson.
\newblock {CPT}-boosted wav2vec2.0: Towards noise robust speech recognition for classroom environments, 2024.

\bibitem{baevski2020wav2vec}
A.~Baevski, Y.~Zhou, A.~Mohamed, and M.~Auli.
\newblock wav2vec 2.0: A framework for self-supervised learning of speech representations.
\newblock {\em Advances in neural information processing systems}, 33:12449--12460, 2020.

\bibitem{pyannote}
H.~Bredin.
\newblock {pyannote.audio 2.1 speaker diarization pipeline: principle, benchmark, and recipe}.
\newblock In {\em Proc. INTERSPEECH 2023}, 2023.

\bibitem{canovas2022analysis}
O.~Canovas and F.~J. García~Clemente.
\newblock Analysis of classroom interaction using speaker diarization and discourse features from audio recordings.
\newblock 09 2022.

\bibitem{cao2023comparative}
J.~Cao, A.~Ganesh, J.~Cai, R.~Southwell, E.~M. Perkoff, M.~Regan, K.~Kann, J.~H. Martin, M.~Palmer, and S.~D'Mello.
\newblock A comparative analysis of automatic speech recognition errors in small group classroom discourse.
\newblock In {\em Proceedings of the 31st ACM Conference on User Modeling, Adaptation and Personalization}, UMAP '23, page 250–262, New York, NY, USA, 2023. Association for Computing Machinery.

\bibitem{chang2020endtoendmultispeakerspeechrecognition}
X.~Chang, W.~Zhang, Y.~Qian, J.~L. Roux, and S.~Watanabe.
\newblock End-to-end multi-speaker speech recognition with transformer, 2020.

\bibitem{demszky2023m}
D.~Demszky and J.~Liu.
\newblock M-powering teachers: Natural language processing powered feedback improves 1: 1 instruction and student outcomes.
\newblock In {\em Proceedings of the Tenth ACM Conference on Learning@ Scale}, pages 59--69, 2023.

\bibitem{demszky2024does}
D.~Demszky, R.~Wang, S.~Geraghty, and C.~Yu.
\newblock Does feedback on talk time increase student engagement? evidence from a randomized controlled trial on a math tutoring platform.
\newblock In {\em Proceedings of the 14th Learning Analytics and Knowledge Conference}, pages 632--644, 2024.

\bibitem{desplanques2020ecapa}
B.~Desplanques, J.~Thienpondt, and K.~Demuynck.
\newblock Ecapa-tdnn: Emphasized channel attention, propagation and aggregation in tdnn based speaker verification.
\newblock {\em arXiv preprint arXiv:2005.07143}, 2020.

\bibitem{dubey2016speaker}
H.~Dubey, L.~Kaushik, A.~Sangwan, and J.~H. Hansen.
\newblock A speaker diarization system for studying peer-led team learning groups.
\newblock {\em arXiv preprint arXiv:1606.07136}, 2016.

\bibitem{dutta-etal-2022-activity}
S.~Dutta, D.~Irvin, J.~Buzhardt, and J.~H. Hansen.
\newblock Activity focused speech recognition of preschool children in early childhood classrooms.
\newblock In E.~Kochmar, J.~Burstein, A.~Horbach, R.~Laarmann-Quante, N.~Madnani, A.~Tack, V.~Yaneva, Z.~Yuan, and T.~Zesch, editors, {\em Proceedings of the 17th Workshop on Innovative Use of NLP for Building Educational Applications (BEA 2022)}, pages 92--100, Seattle, Washington, July 2022. Association for Computational Linguistics.

\bibitem{fujita2020neural}
Y.~Fujita, S.~Watanabe, S.~Horiguchi, Y.~Xue, J.~Shi, and K.~Nagamatsu.
\newblock Neural speaker diarization with speaker-wise chain rule.
\newblock {\em arXiv preprint arXiv:2006.01796}, 2020.

\bibitem{fung2016collaborative}
D.~Fung, H.~To, and K.~Leung.
\newblock collaborative-learning: The influence of collaborative group work on students’ development of critical thinking: The teacher’s role in facilitating group discussions.
\newblock {\em Pedagogies: An International Journal}, 11:146--166, 04 2016.

\bibitem{gomez2022speakerdiarizationidentificationsinglechannel}
A.~Gomez.
\newblock Speaker diarization and identification from single-channel classroom audio recording using virtual microphones, 2022.

\bibitem{GoogleCloud2021}
{Google Cloud}.
\newblock Detect different speakers in an audio recording, 2021.
\newblock \url{https://cloud.google.com/speech-to-text/docs/multiple-voices}.

\bibitem{harper2019nemo}
E.~Harper, S.~Majumdar, O.~Kuchaiev, L.~Jason, Y.~Zhang, E.~Bakhturina, V.~Noroozi, S.~Subramanian, K.~Nithin, H.~Jocelyn, et~al.
\newblock Nemo: A toolkit for conversational ai and large language models.
\newblock {\em Computer software], URL: https://github. com/NVIDIA/NeMo}, 2019.

\bibitem{w2v2_robust_large}
W.-N. Hsu, A.~Sriram, A.~Baevski, T.~Likhomanenko, Q.~Xu, V.~Pratap, J.~Kahn, A.~Lee, R.~Collobert, G.~Synnaeve, et~al.
\newblock Robust wav2vec 2.0: Analyzing domain shift in self-supervised pre-training.
\newblock {\em arXiv preprint arXiv:2104.01027}, 2021.

\bibitem{kanda2022transcribe}
N.~Kanda, X.~Xiao, Y.~Gaur, X.~Wang, Z.~Meng, Z.~Chen, and T.~Yoshioka.
\newblock Transcribe-to-diarize: Neural speaker diarization for unlimited number of speakers using end-to-end speaker-attributed asr.
\newblock In {\em ICASSP 2022-2022 IEEE International Conference on Acoustics, Speech and Signal Processing (ICASSP)}, pages 8082--8086. IEEE, 2022.

\bibitem{kelly2018automatically}
S.~Kelly, A.~Olney, P.~Donnelly, M.~Nystrand, and S.~D’Mello.
\newblock Automatically measuring question authenticity in real-world classrooms.
\newblock {\em Educational Researcher}, 47:0013189X1878561, 06 2018.

\bibitem{classbank}
B.~MacWhinney.
\newblock A transcript-video database for collaborative commentary in the learning sciences.
\newblock In R.~Goldman, R.~Pea, B.~Barron, and S.~Derry, editors, {\em Video research in the learning sciences}, pages 537--546. Lawrence Erlbaum Associates, Mahwah, NJ, 2007.

\bibitem{NemoECAPATDNN2024}
NVIDIA.
\newblock Ecapa-tdnn model for speaker recognition and verification, 2024.
\newblock Accessed: February 15, 2025.

\bibitem{NemoDiarization2024}
NVIDIA.
\newblock {\em NVIDIA NeMo Speaker Diarization User Guide}, 2024.
\newblock Accessed: February 15, 2025.

\bibitem{NemoSpeakerNet2024}
NVIDIA.
\newblock Speakernet model for speaker recognition and verification, 2024.
\newblock Accessed: February 15, 2025.

\bibitem{NemoTitaNetLarge2024}
NVIDIA.
\newblock Titanet-large model for speaker recognition and verification, 2024.
\newblock Accessed: February 15, 2025.

\bibitem{vad_marblenet}
NVIDIA.
\newblock Vad marblenet model, 2024.
\newblock Accessed: February 15, 2025.

\bibitem{vad_frame_marblenet}
NVIDIA.
\newblock Vad multilingual frame marblenet model, 2024.
\newblock Accessed: February 15, 2025.

\bibitem{vad_multilingual_marblenet}
NVIDIA.
\newblock Vad multilingual marblenet model, 2024.
\newblock Accessed: February 15, 2025.

\bibitem{vad_telephony_marblenet}
NVIDIA.
\newblock Vad telephony marblenet model, 2024.
\newblock Accessed: February 15, 2025.

\bibitem{Park_2020}
D.~S. Park, Y.~Zhang, Y.~Jia, W.~Han, C.-C. Chiu, B.~Li, Y.~Wu, and Q.~V. Le.
\newblock Improved noisy student training for automatic speech recognition.
\newblock {\em arXiv preprint arXiv:2005.09629}, 2020.

\bibitem{park2021reviewspeakerdiarizationrecent}
T.~J. Park, N.~Kanda, D.~Dimitriadis, K.~J. Han, S.~Watanabe, and S.~Narayanan.
\newblock A review of speaker diarization: Recent advances with deep learning, 2021.

\bibitem{radford2022robustspeechrecognitionlargescale}
A.~Radford, J.~W. Kim, T.~Xu, G.~Brockman, C.~McLeavey, and I.~Sutskever.
\newblock Robust speech recognition via large-scale weak supervision, 2022.

\bibitem{whisper}
A.~Radford, J.~W. Kim, T.~Xu, G.~Brockman, C.~McLeavey, and I.~Sutskever.
\newblock Robust speech recognition via large-scale weak supervision.
\newblock In {\em International conference on machine learning}, pages 28492--28518. PMLR, 2023.

\bibitem{rasanen2021alice}
O.~R{\"a}s{\"a}nen, S.~Seshadri, M.~Lavechin, A.~Cristia, and M.~Casillas.
\newblock Alice: An open-source tool for automatic measurement of phoneme, syllable, and word counts from child-centered daylong recordings.
\newblock {\em Behavior Research Methods}, 53:818--835, 2021.

\bibitem{tim_sainburg_2019_3243139}
T.~Sainburg.
\newblock timsainb/noisereduce: v1.0, June 2019.

\bibitem{sainburg2020finding}
T.~Sainburg, M.~Thielk, and T.~Q. Gentner.
\newblock Finding, visualizing, and quantifying latent structure across diverse animal vocal repertoires.
\newblock {\em PLoS computational biology}, 16(10):e1008228, 2020.

\bibitem{southwell2022challenges}
R.~Southwell, S.~Pugh, E.~M. Perkoff, C.~Clevenger, J.~Bush, R.~Lieber, W.~Ward, P.~Foltz, and S.~D'Mello.
\newblock Challenges and feasibility of automatic speech recognition for modeling student collaborative discourse in classrooms.
\newblock In A.~Mitrovic and N.~Bosch, editors, {\em Proceedings of the 15th International Conference on Educational Data Mining}, pages 302--315, Durham, United Kingdom, July 2022. International Educational Data Mining Society.

\bibitem{sun2024said}
A.~Sun, J.~J. Londono, B.~Elbaum, L.~Estrada, R.~J. Lazo, L.~Vitale, H.~G. Villasanti, R.~Fusaroli, L.~K. Perry, and D.~S. Messinger.
\newblock Who said what? an automated approach to analyzing speech in preschool classrooms.
\newblock {\em arXiv preprint arXiv:2401.07342}, 2024.

\bibitem{suresh2021using}
A.~Suresh, J.~Jacobs, V.~Lai, C.~Tan, W.~Ward, J.~H. Martin, and T.~Sumner.
\newblock Using transformers to provide teachers with personalized feedback on their classroom discourse: The talkmoves application.
\newblock {\em arXiv preprint arXiv:2105.07949}, 2021.

\bibitem{wang2024speaker}
J.~Wang, S.~Dudy, X.~He, Z.~Wang, R.~Southwell, and J.~Whitehill.
\newblock Speaker diarization in the classroom: How much does each student speak in group discussions?
\newblock In B.~PaaÃŸen and C.~D. Epp, editors, {\em Proceedings of the 17th International Conference on Educational Data Mining}, pages 360--367, Atlanta, Georgia, USA, July 2024. International Educational Data Mining Society.

\bibitem{wang2014automatic}
Z.~Wang, X.~Pan, K.~F. Miller, and K.~S. Cortina.
\newblock Automatic classification of activities in classroom discourse.
\newblock {\em Computers \& Education}, 78:115--123, 2014.

\bibitem{denoising3_wung}
J.~Wung, B.-H.~F. Juang, and B.~Lee.
\newblock Speech enhancement based on a log-spectral amplitude estimator and a postfilter derived from clean speech codebook.
\newblock In {\em 2010 18th European Signal Processing Conference}, pages 999--1003, 2010.

\bibitem{denoising1_yu}
G.~Yu, E.~Bacry, and S.~Mallat.
\newblock Audio signal denoising with complex wavelets and adaptive block attenuation.
\newblock In {\em 2007 IEEE International Conference on Acoustics, Speech and Signal Processing - ICASSP '07}, volume~3, pages III--869--III--872, 2007.

\bibitem{denoising2_zhou}
J.~Zhou and L.~Tao.
\newblock Speech enhancement in joint time-frequency domain based on real-valued discrete gabor transform.
\newblock In {\em 2010 5th International Conference on Computer Science \& Education}, pages 1028--1031, 2010.

\end{thebibliography}
\end{document}